\documentclass{article}
\usepackage{spconf,amsmath,graphicx}
\usepackage{lipsum}

\title{5G Routing Interfered Environment}
\name{Barak Gahtan\thanks{barakgahtan@cs.technion.ac.il}}
\address{Technion}

\begin{document}
%
\maketitle
%

\section{ABSTRACT}\label{sec:intro}
5G is the next-generation cellular network technology, with the goal of meeting the critical demand for bandwidth required to accommodate a high density of users. It employs flexible architectures to accommodate the high density \cite{7390965}. 5G is enabled by mmWave communication, which operates at frequencies ranging from 30 to 300 GHz. This paper discusses the creation of a python-based environment known as the 5G Routing Interfered Environment (5GRIE). The environment can run different algorithms to route packets with source and destination pairs using a formulated interference model. Deep Reinforcement Learning algorithms that use Stable-Baselines 3 \cite{Raffin_Stable_Baselines3_2020} based on Gym, as well as heuristic-based algorithms like random or greedy, can be run on it. Profitable is an algorithm that is provided.

\section{Interference}\label{sec:interference}
The interference model used in 5GRIE is as follows. Let $S_i$ denote station $i$ in the topology. In addition, let $l=(S_i \rightarrow S_j)$ denote the mmWave link used by $S_i$ for transmitting packets to $S_j$. Let the power used by this link be $P(l)$. The received power at $S_j$ is given by 
\begin{equation}
\begin{gathered}
P_\mathrm{Received}(l) = P(l) - \mathrm{FSL(S_i,S_j)}-\eta,
\end{gathered}
\nonumber
\label{eq:22}
\end{equation}
where the free space path loss is $\mathrm{FSL(S_i,S_j)} = \left(\frac{c}{4 \pi  D(S_i,S_j) f}\right)^{2}.$
\textit{c} is the speed of light, $D(S_i,S_j)$ is the distance between stations $S_i$ and $S_j$, $f$ is the radio carrier frequency and $\eta$ is a uniform noise that simulates bad weather conditions. Let $l' = (S_k \rightarrow S_n)$ be another link transmitting at power $P(l')$. This link creates an interference at $S_j$, whose power is given by $P(l')\cdot I(l,l')$, where
\begin{equation}
\begin{gathered}
I(l,l') = \mathrm{AngleToPower}(\alpha(l,l'))(P(l') - \mathrm{FSL(S_i',S_j)} - \eta).
\end{gathered}
\nonumber
\end{equation} 
Here, 
\begin{equation}
\mathrm{AngleToPower}(\alpha) = 
\begin{cases}
    1-\frac{\alpha}{90}  & 0 \leq \alpha \leq 90^{\circ}\\
    0               & \text{otherwise},
\end{cases}
\nonumber
\label{eq:1}
\end{equation} simulating the main and side lobes of the angle $\alpha$, modeled as a triangle, it 
is the directivity of $S_j$'s antenna. $\alpha(l,l')$ is the angle between $S_i,\ S_j$ and $S_k$. Assuming all interferences are add up, we define the effective receive power of link $l$ as
$P_\mathrm{Received_\mathrm{eff}}(l) =  P_\mathrm{Received}(l)  - \sum_{l' \ne l} I(l,l').$
Thus, the actual capacity of link $l$ is given by
\begin{equation}
C(l) = \frac{P_\mathrm{Received_\mathrm{eff}}(l)}{P_\mathrm{Received}(l)} C_\mathrm{nom}(l),
\label{equationpositive}
\end{equation}
where $C_\mathrm{nom}(l)$ is the link nominal capacity, in the absence of any interference. The number of packets that can be transmitted per step, over link $l$ is given by $N_l = C_l \Delta t$, where $\Delta t$ is the time step duration.

\section{Profitable Algorithm}\label{sec:greedy}
The profitable algorithm (Profitable) keeps a set $L$ of all the mmWave links it's considering for the next step, as well as a set $DL$ of all the mmWave links it's already made a decision on. At first, $L$ contains all of the links, whereas $DL$ is empty. During each iteration, Profitable considers one link $l$, which is chosen at random from $L$. Then it determines whether or not to add $l$ to $DL$, as well as how much power to use with that link.

Adding a link $l$ to $DL$ has both positive and negative effects on network performance. 
\begin{description}
\item[Positive Part] - $C_+$, comes from the fact that packets can be sent over this link.
\item[Negative Part] Interference between the new link and previously selected links causes the negative part, $C_-$. 
\end{description}
The difference between $C_+$ and $C_-$ is defined as the \textit{profit}.

Eq. \ref{equationpositive} gives the capacity profit of adding $l$ into $DL$ using power $P_\mathrm{Received}$, which indicates the number of packets that will be transmitted over this link during the next step if it is activated using the tested power. 

Profitable then computes the difference between the new and old effective received power of $l'$ while considering the new link $l$ and its $P_\mathrm{Received}$ for each link $l'$ previously added to $DL$. The number of packets that could not be routed in the next step due to link $l$ interference is the capacity loss:
\begin{equation}
\begin{gathered}
C_- = \sum_{l'  \in \mathrm{S}}{(P_\mathrm{Received_\mathrm{eff}}(l'_{old})-P_\mathrm{Received_\mathrm{eff}}(l'_{new}))}\frac{C_\mathrm{nom}(l')}{P_\mathrm{Received}(l')}.
\nonumber
\end{gathered}
\end{equation}

To summarize, Profitable selects a random link $l$ from $L$, and performs the following actions until $L$ is empty:
\begin{enumerate}
    \item It calculates the \textit{profit} of adding $l$ with power $p$ for each possible power.
    \item If the \textit{profit} is positive for some $p$, $l$ is added to $DL$ with the $p$ that has the highest \textit{profit}. else, $l$ is not added to $DL$.
\end{enumerate} When the iterations end, $DL$ contains the selected power for the next step for each mmWave link when $L$ is empty.\\
\section{Environment Implementation}\label{sec:env}
The environment includes five models: packets, buffers, stations, network topology, and the Interference Model. We will now go over each model in greater detail.
\begin{table}[tb]
\small
\centering
\scalebox{0.9}{
\begin{tabular}{l|| cl c l c}
	\hline \hline
	Field & Purpose   \\
    \hline
    Source & Where did the packet originate from. \\
    Destination & Destination of the packet. \\
    Current location & Current location of the packet. \\
    Shortest path list & List of the shortest path of the packet.  \\
    Next hop & Next hop station of the packet. \\
    Number of packets & Number of packets. \\
    \hline \hline
\end{tabular}}
\caption{Packet fields and purpose.}
\label{table:packet}
\end{table}

\subsection{Packets}
The packets fields are listed in Table \ref{table:packet}. This model contains several methods.
\begin{description}
    \item[packet step] Changes the packet's current location to the next hop, if one exists. If it reaches its destination, it returns the number of packets received.
    \item[get next hop]Returns the next hop station of the packet.
\end{description}

\subsection{Buffers}
\begin{table}[tb]
\small
\centering
\scalebox{0.9}{
\begin{tabular}{l|| cl c l c}
	\hline \hline
	Field & Purpose   \\
    \hline
    Name & ID of the station. \\
    Connection matrix & Information of the topology. \\
    Shortest path list & List of the shortest path in the topology. \\
    Out links & Dictionary of mmWave out-going links.  \\
    Max transceiver & Maximum number of transceivers. \\
    Current transceiver & Current number of transceivers. \\
    \hline \hline
\end{tabular}}
\caption{Station's fields and purpose.}
\label{table:station}
\end{table}
\begin{table}[tb]
\small
\centering
\scalebox{0.7}{
\begin{tabular}{l|| cl c l c}
	\hline \hline
	Field & Purpose   \\
    \hline
    Source name & Name of the station the buffer is located in. \\
    Out going link to & Name of which station the queue is intending the send packets to.\\
    Flows & Dictionary containing packets that are waiting in the buffer. \\
    Total flows & The number of different flows that the buffer currently holds. \\
    Link's maximum capacity & The associated mmWave's link maximum capacity per step.\\
    Used bw & Amount of used bandwidth of the current step.  \\
    Power & The chosen level of power used by the mmWave link   \\
     & associated with a source and destination of the buffer. \\
    Current packets & Total number of packets in the buffer. \\
    Max packets & Maximum number of packets the buffer can hold, before dropping packets.  \\
    Dropped packets & The number of dropped packets this buffer has dropped during a step.\\ 
    \hline \hline
\end{tabular}}
\caption{Buffer's fields and purpose.}
\label{table:buffer}
\end{table}

The buffer's fields are listed in Table \ref{table:buffer}. This model contains several methods.
\begin{description}
    \item[add flow to q] - Adds a packet-based flow to the buffer. If the buffer overflows, packets are dropped.
    \item[remove flow from q] - Removes a packet-based flow from the buffer. Additionally, the relevant fields are updated.
    \item[zero bw in buffer] - The used band width and dropped packets fields are set to zero.
    \item[get total data] - Returns the number of total packets that are in the buffer.
    \item[get dropped packets in q] - The total number of packets dropped during the step is returned.
\end{description}

\subsection{Station}

The station's fields are listed in Table \ref{table:station}. This model contains several methods.
\begin{description}
    \item[initialize out queues] - Sets the buffers in that station to their default values.
    \item[add flow] - Adds a flow into the station's buffers. 
    \item[is link activated] - If the link is active for the current step, this function returns true.
    \item[remove flow] - This function removes a flow from the station's buffer.
    \item[update active links] - The method activates the mmWave links with the specified power levels for each of them.
    \item[get buffers observations] - Returns a dictionary containing a triple - total number of data packets in the buffer, percentage load on each buffer - between [0,1] and total number of dropped packets by the buffer during the step - for each buffer in the station.
    \item[zero bw] - Method zero bw in buffer is called for each of the station's buffers.
    \item[get dropped packets] - Returns the sum of dropped packets for each of the station's buffers by calling "get dropped packets in q."
\end{description}

\subsection{Interference Model}
\ref{sec:interference} explains the environment's interference model. Each instance of the interference model is a $I(l,l')$ matrix of size [number of mmWave links x number of mmWave links], with $l$ interfering all other mmWave links $l'$ by the value in the matrix.

\subsection{Network Topology}
The task of creating a topology. Python's networkx module is used to generate the topology. The input is a dictionary of dictionaries containing the link weights for mmWave. A uniform distribution of weights is used to select each weight at random.

We will now describe the environment in which the Deep Reinforcement Learning algorithm will operate, using Stable Baselines3 agents \cite{Raffin_Stable_Baselines3_2020}. 
Table \ref{table:env} lists the environmental fields. The main methods for the environment:
\begin{description}
    \item[generate demand random] Generates random data for the next episode. SB3's regular reset method makes use of this. It generates (demand matrix, total packets, flow list, and interference) for the following episode.
    \item[process flows] Moves packets in the system one step. It decreases the total packet counter until it reaches zero. An episode ends when the number of flows in the system reaches zero.
    \item[get dropped packets] Returns the sum of dropped packets for the passing step.
    \item[get state observation] Returns the next state's observation. 
    \item[adopt interference] Before proceeding to the next step, it computes the effective power of each mmWave link using the interference matrix.
    \item[update active links and bw] Before the start of the step, the method first zeroes the various mmWave links' used bandwidth in the system, and then it calls "update interfernce" to update the next effective power of each mmWave link in the system, according to the chosen actions. 
    \item[reset] The standard reset method for SB3, inheriting from Gym. For the training phase, this reset generates data at random.
    \item[reset custom] A special reset method for evaluating DRL algorithms against the heuristics-based algorithm. This reset uses data from a list that was created at the start of the program's execution, ensuring that the comparison is exact on the same data.
    \item[reward] Reward funtion for SB3 to use. 
    \item[step] This is the step in which the SB3 library is put to use,  inheriting from Gym. The environment employs all models and moves packets until they are dropped or reach their final destination at each step.
    \item[convert actions to edges] It is an adopter method in two cases, depending on whether a hueristics-based or DRL algorithm is used. It associates an action with a specific mmWave link in the topology; for example, if the vector's i'th entry is $0.5$, it means that when id=i is translated to edges using the dictionary saved, the specific mmWave link will use a power level of $0.5$.
\end{description}

\begin{table}[tb]
\small

\centering
\scalebox{0.7}{
\begin{tabular}{l|| cl c l c}
	\hline \hline
	Field & Purpose   \\
    \hline
    Edges to ID & Mapping mmWave links to IDs. \\
    ID to edges & Mapping IDs to mmWave links.\\
    Step count & Number of steps of the current episode. \\
    Net & Topology. \\
    Episode count & Number of episodes.\\
    Dropped packets & Number of dropped packets for the episode.  \\
    Global all shortest path & List of list of shortest paths of the topology.  \\
    Edges info & Panda's frame of mmWave links with their weights. \\
    Interference  & The interference model being currently used.  \\
    Next episode demand matix & The demand matrix for the next episode.\\ 
    Routers list & Dictionary of stations.\\ 
    Episodes demand eval & List of data to compare with heuristics based algorithms.\\ 
     & Data is (demand matrix, total packets, list of flows and interference) \\ 
    Episodes demand train & Same as above, but for verbose and training. \\
    Observation space & Observation space that will be used by SB3. \\
    Action space & Action space that will be used by SB3. \\
    \hline \hline
\end{tabular}}
\caption{Environment's fields and purpose.}
\label{table:env}
\end{table}

\section{Conclusion}\label{sec:conclusion}
We describe a Python-based implementation of an environment - 5GRIE - that allows the use of DRL agents implemented in the SB3 library and hueristics-based algorithms in this paper. To our knowledge, this is the first time a simulator of 5G packet routing between source and destination pairs has been implemented using a formulated interference model. Additional requirements for running and using the environment are listed in the READ-ME file.\\
The code for our environment can be found in the repository:
https://github.com/BarakGahtan/5GIRE.
\bibliographystyle{IEEEbib}
\bibliography{refs}

\end{document}